\documentstyle[12pt]{article}
\begin{document}
\input epsf
\begin{titlepage}
\begin{center}
\today     \hfill    WM-98-113\\
~{} \hfill hep-ph/9809522 \\

\vskip .5in

{\large \bf Limits on a Light Leptophobic Gauge Boson}

\vskip 0.3in

Alfredo Aranda\footnote{fefo@physics.wm.edu} 
and Christopher D. Carone\footnote{carone@physics.wm.edu}

\vskip 0.1in

{\em Nuclear and Particle Theory Group \\
     Department of Physics \\
     College of William and Mary \\
     Williamsburg, VA 23187-8795}


        
\end{center}

\vskip .1in

\begin{abstract}
We consider the phenomenology of a naturally leptophobic $Z$-prime boson 
in the $1$ to $10$ GeV mass range.  The $Z$-prime's couplings to leptons 
arise only via a radiatively-generated kinetic mixing with the $Z$ and photon,
and hence are suppressed.  We map out the allowed regions of the 
mass-coupling plane and show that such a $Z$-prime that couples to quarks 
with electromagnetic strength is not excluded by the current data.  We 
then discuss possible signatures at bottom and charm factories.
\end{abstract}

\end{titlepage}

\newpage
\renewcommand{\thepage}{\arabic{page}}
\setcounter{page}{1}
\section{Introduction} \label{sec:intro} \setcounter{equation}{0}

In the past few years, the possibility of new leptophobic gauge bosons
has been explored as a means of explaining apparent discrepancies in
electroweak observables measured with high precision 
at LEP \cite{eo1,eo2}, as well as an apparent high $E_T$ excess in 
the inclusive dijet spectrum at the Tevatron \cite{et}.   While for the 
most part these anomalies have since gone away, the possibility remains 
that a $Z$-prime boson  ($Z'$) coupling mostly to quarks and with a mass 
smaller than $m_Z$ could exist while evading experimental 
detection \cite{carmur1,carmur2,db}.  Given the assumptions
that (1) the leptons are not charged under the new U(1) gauge interaction,
and (2) the couplings to quarks are generation independent (to
avoid large flavor-changing neutral current effects) then the normalization 
of the U(1) can be chosen so that the $Z'$ couples precisely to baryon 
number.   Anomaly cancellation can be achieved at the expense of
introducing new exotic states.  Two explicit examples of viable,
anomaly-free models were presented in Refs.~\cite{carmur1,carmur2}, and 
these models presumably don't exhaust the possible ways in which 
anomalies can be cancelled.  Therefore, we will set model-building issues
aside and focus instead on the phenomenology of the $Z'$.  This is of 
timely interest given the recent stringy suggestion that the Planck scale and
weak scale might be identified \cite{nima,ddg}.  In these scenarios, the 
dimension-5 baryon and lepton number violating operators that arise 
generically at the string scale would only be suppressed by a TeV, and 
hence would be phenomenologically lethal.  Barring a higher-dimensional 
solution to the proton decay problem \cite{ddg}, additional gauge symmetries 
could provide a more prosaic, though equally effective, resolution.

In Ref.~\cite{carmur2}, a specific mechanism was proposed for maintaining
leptophobia in models with gauged baryon number, and we will adopt this
mechanism here.  The reason that leptophobia is not automatic is
that the baryon number and hypercharge gauge fields mix via their kinetic
terms
\begin{equation}
{\cal L}_{kin}= -\frac{1}{4} (F^{\mu \nu}_Y F^Y_{\mu \nu} +
                   2c F^{\mu \nu}_B F^Y_{\mu \nu} + F^{\mu \nu}_B 
                   F^B_{\mu \nu}) \,\,\, .
\end{equation}
We assume there are no Higgs fields that carry both baryon number
and electroweak quantum numbers, so that mass mixing terms are not 
present.  Below the electroweak symmetry breaking scale, there are 
separate kinetic mixing parameters for the photon and $Z$, which we will call
$c_\gamma$ and $c_Z$, respectively.  In order that leptophobia be preserved,
$c_\gamma$ and $c_Z$ must be sufficiently small at experimental energies.  
This can be arranged if the parameter $c$ is forced to zero at some high 
scale $\Lambda$, so that $c_\gamma$ and $c_Z$ are only generated 
at the one-loop level, via renormalization group running.  The
boundry condidion $c(\Lambda)=0$ can be achieved, for example, by
embedding U(1)$_B$ into a non-Abelian group, as was shown explicitly in 
Ref.~\cite{carmur2}.  Here we will be more general and not assume
the specific mechanism for achieving this boundary condition.  Thus, the
boundary condition, together with assumptions (1) and (2) given above, 
define a class of models that we will consider further in the present 
analysis.

In Ref.~\cite{carmur2}, the $Z'$ mass range $m_\Upsilon < m_B < m_Z$ 
was studied, primarily because the coupling $\alpha_B$ could be
taken as large as $\sim 0.1$ at points within this interval, without 
conflicting with the experimental bounds.  Possible high energy collider
signatures were then considered.  Here we will focus instead on 
$Z'$ masses between $\sim 1$ GeV and $m_\Upsilon$, with the
initial goal of determining  how tightly we can bound the parameter space 
of the model.   Although the coupling $\alpha_B$ cannot be as large as
$0.1$ within this mass range, we will show that current experiment
does allow it to be comparable to $\alpha_{EM} \approx 1/137$.  Given this
result, we consider the possibility of detecting the $Z'$ at charm and bottom 
meson factories via the decays of various quarkonium states which would be 
plentifully produced.  We will not consider smaller values of $m_B$, but 
instead refer the interested reader to the discussion in Ref~\cite{neltet}.

This letter is organized in two parts.  We will first discuss the current
bounds on the parameter space of the model. With the boundary condition on 
the kinetic mixing terms described above, both the hadronic and leptonic 
signatures of the $Z'$ are completely determined by its mass, $m_B$, 
and gauge coupling $g_B = \sqrt{4\pi\alpha_B}$.  Therefore, these bounds
can be translated into  boundaries of excluded regions on a two-dimensional 
mass-coupling plane.  We will then consider possible discovery signals 
for a $Z'$ living within these allowed regions.

\section{Parameter Space}

Most of the important phenomenological bounds follow directly from the 
$Z$-prime's gauge coupling to quarks.  In addition, we take into account
the small kinetic mixing effects by treating the mixing term in 
$\mathcal{L}$$_{kin}$ as a perturbative interaction.  The Feynman rules 
corresponding to the $Z^{'}-\gamma$ and $Z^{'}-Z$ vertices are
\begin{eqnarray}
-ic_{\gamma}\cos\theta_{w}(p^2g^{\mu\nu}-p^{\mu}p^{\nu}) ,
\end{eqnarray}
and
\begin{eqnarray}
ic_{Z}\sin\theta_{w}(p^2g^{\mu\nu}-p^{\mu}p^{\nu}) ,
\end{eqnarray} 
respectively, where $c_{\gamma}=c_{Z}=c$ above the electroweak scale,
and where $c=0$ at some ultraviolet cutoff $\Lambda$.  We will 
initially set $\Lambda=m_{top}\approx 180$~GeV, since this is 
probably the lowest scale at which the new physics responsible for the 
boundary condition $c(\Lambda)=0$ might itself remain undetected.  We 
will describe how our results change with different choices for $\Lambda$
as needed.  Note that choosing a somewhat higher value for $\Lambda$, for
example $500$~GeV, has only a small effect on the mixing since the 
dependence on $\Lambda$ is only logarithmic. 

At any desired renormalization scale $\mu$, we may rewrite $c_\gamma(\mu)$ and 
$c_Z(\mu)$ as an explicit function of $\alpha_B$ by solving the one-loop 
renormalization group equations.  These equations follow from the one 
quark-loop diagrams that connects the $Z^{'}$ to the $\gamma$ and $Z$, 
respectively \cite{carmur2}:
\begin{eqnarray}
\mu\frac{\partial}{\partial\mu}c_{\gamma}(\mu)=-\frac{2}{9\pi}\frac{
\sqrt{\alpha_{B}\alpha}}{c_{w}}[2N_{u}-N_{d}]
\end{eqnarray}
and
\begin{equation}\label{cz}
\mu\frac{\partial}{\partial\mu}c_{Z}(\mu)=-\frac{1}{18\pi}\frac{
\sqrt{\alpha_{B}\alpha}}{s_{w}^2c_{w}}[3(N_{d}-N_{u})+4(2N_{u}-N_{d})s_{w}^2]
\,\,\, .
\end{equation}
Here $c_w$ ($s_w$) represents the cosine (sine) of the weak mixing angle,
$\alpha$ is the electromagnetic fine structure constant, and $N_u$ ($N_d$)
is the numbers of charge $2/3$ ($-1/3$) quarks that are lighter than the 
renormalization scale.  It is straightforward to show, for example
\[
c_\gamma (m_b) = 0.033 \sqrt{\alpha_B} \,\,\,\,\,\,\,\, 
c_Z(m_b) = 0.116 \sqrt{\alpha_B}
\]\begin{equation}
c_\gamma (m_c) = 0.047 \sqrt{\alpha_B} \,\,\,\,\,\,\,\,
c_Z(m_c) = 0.130 \sqrt{\alpha_B}
\label{eq:cees}
\end{equation}
We will use expressions like these to translate bounds on leptonic
processes to exclusion regions on the $m_B$-$\alpha_B$ plane.

\begin{figure}  

\centerline{ \epsfxsize=4.5 in \epsfbox{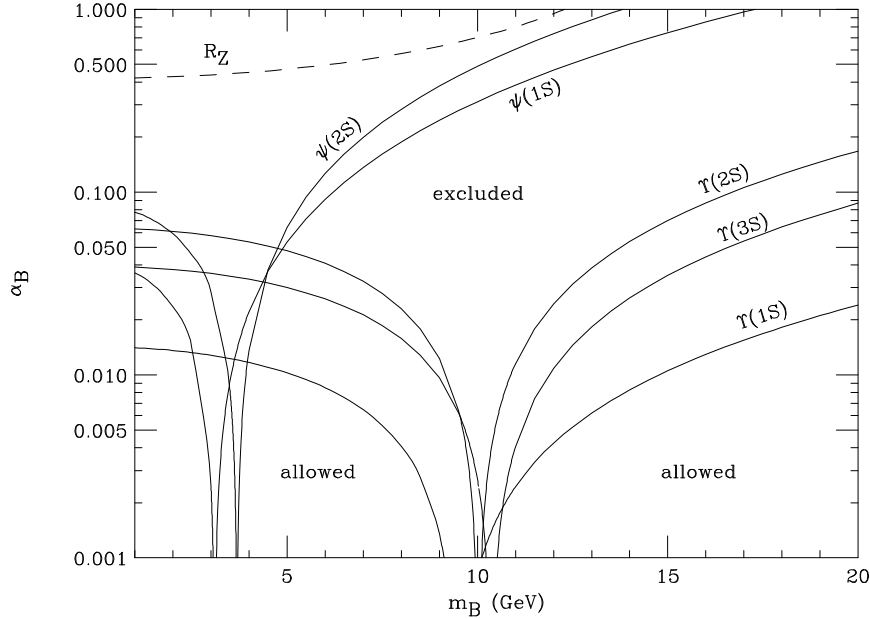}  }

\vglue .3 in

\caption{Bounds from hadronic decays.}

\label{figone}

\end{figure}

The experimental bounds on the model from hadronic decays are summarized
in Fig.~\ref{figone}.  Beginning with the $\Upsilon (1S)$, the new
contribution to the hadronic decay width is given by \cite{carmur1}
\begin{equation}
\Delta R_\Upsilon=\frac{4}{3} \left[ \frac{\alpha_B}{\alpha}
\frac{m^2_\Upsilon}{m_B^2-m_\Upsilon^2} + 
\left(\frac{\alpha_B}{\alpha}\frac{m^2_\Upsilon}{m_B^2-m_\Upsilon^2}
\right)^2\right] \,\,\, ,
\label{eq:rup}
\end{equation}
where $R_\Upsilon \equiv \Gamma (\Upsilon \rightarrow \mbox{hadrons})/
\Gamma (\Upsilon \rightarrow \mu^+ \mu^-)$, and the interference with
$s$-channel photon exchange is included.  The most stringent bound on 
this quantity follows from an ARGUS limit on the non-electromagnetic (NE) 
contribution to the $\Upsilon (1S) \rightarrow 2$~jets branching fraction
\cite{argus},  
\[
BF(\Upsilon(1S)\rightarrow \mbox{2 jets}, \mbox{ NE}) < 0.053 
\,\,\,\, (95\% \mbox{ CL}) \,\,\, ,
\]
which we find corresponds to $\Delta R_\Upsilon < 2.48$.  This bound is 
stronger than the one obtained from the $\Upsilon (1S)$ hadronic width, 
discussed in Ref.~\cite{carmur1}.   Note that we have chosen to restrict
Fig.~\ref{figone} to values of the coupling $\alpha_B > 10^{-3}$, where
direct experimental detection of the $Z'$ via rare decays might be 
feasible.  With this choice, finite width effects omitted from 
Eq.~(\ref{eq:rup}) have a negligible effect on the segments of the 
exclusion curves shown.

We may place additional bounds on the parameter space of the model by 
considering the hadronic decay widths of the $\Upsilon (2S)$ and 
$\Upsilon (3S)$ respectively.  Since no direct experimental bounds exist 
on the non-electromagnetic, two jet branching fraction, we compare 
$R_{\Upsilon (2S)}$ and $R_{\Upsilon (3S)}$ to the perturbative QCD 
prediction \cite{mac},
\begin{equation}
R  = \frac{10(\pi^2-9)}{9\pi}\frac{\alpha_s^3}{\alpha^2}\left(
1+\frac{\alpha_s}{\pi}\left\{-18.2 + \frac{3}{2}\beta_0 [
1.161 + \ln (\frac{2\mu}{m_\Upsilon})] \right\}\right)
\label{eq:3glu}
\end{equation}
where $\beta_0=11-2 n_f/3 = 23/3$.  We evaluate this expression using
$\alpha_s(m_b)$ as determined from the world average value 
$\alpha_s(m_Z) = 0.119 \pm 0.002$ \cite{rpp}.  We extract the experimental
values of $R$ from branching fraction data in the 1998 Review of Particle 
Properties \cite{rpp}.  This is straightforward, except in the
case of the $\Upsilon (3S)$, where the branching fraction
to $\mu^+\mu^-$ has not been measured.  We assume in this case 
that $\Gamma(\mu^+\mu^-)$ is approximately equal to $\Gamma(e^+e^-)$, 
which has been measured.  Taking into account experimental uncertainties, 
we find $\Delta R \mbox{\raisebox{-1.0ex} {$\stackrel{\textstyle ~<~}
{\textstyle \sim}$}} 92$ and $\Delta R  \mbox{\raisebox{-1.0ex} 
{$\stackrel{\textstyle ~<~} {\textstyle \sim}$}}  33$ for the 
$\Upsilon (2S)$ and $\Upsilon (3S)$ respectively, at the 95\% confidence
level.  Although these bounds are weak, they nonetheless exclude 
some additional region of the parameter space immediately around the 
resonance masses.

Similar bounds may be determined from the hadronic decay widths of 
the $J/\psi$ and the $\psi (2S)$.  Here, however, it is not so
straigtforward to determine the standard model expectation.
The perturbative QCD prediction for the gluonic decay width in 
Eq.~(\ref{eq:3glu}) is derived in a nonrelativistic bound state 
approximation, and is therefore subject to O($v^2/c^2$) corrections, which
are expected to be significant. Therefore, we will use the results of 
a recent relativistic potential model analysis \cite{pma} as as our 
standard model expectation.  In Ref.~\cite{pma}, the $J/\psi$ hadronic 
decay width was used to extract $\alpha_s (m_c)$, yielding $0.29 \pm 0.02$.  
Comparing to the world average value, we find that the difference 
$\Delta \alpha_s (m_c) < 0.068$ can be tolerated, allowing two-standard 
deviation uncertainties.  Thus, any new contribution to $R_\psi$ is bounded by 
$\Delta R \mbox{\raisebox{-1.0ex} {$\stackrel{\textstyle ~<~} 
{\textstyle \sim}$}} 3 (\Delta\alpha_s/\alpha_s) R \approx 34$,
yielding  the contour shown in Fig.~\ref{figone}.  We detemine the
gluonic contribution to $R_{\psi (2S)}$ from branching fraction data 
in the Review of Particle Physics \cite{rpp}, and obtain
$R_g = 123.6 \pm 27.3$.  Since this is so large and uncertain, the
bound on the model's parameter space will clearly be weak.  Thus, we simply
compare $R_{\psi (2S)}$ to the perturbative QCD prediction, including a 
theoretical uncertainty comparable in size to the relativistic corrections 
in the $J/\psi$ case; we find 
$\Delta R \mbox{\raisebox{-1.0ex} {$\stackrel{\textstyle ~<~} 
{\textstyle \sim}$}} 162$, yielding the curve shown.  

Finally, Fig.~\ref{figone} displays the bound from the hadronic decay
width of the $Z$, labelled $R_Z$, which we find provides the strongest 
constraint from the $Z$-pole observables.  This result includes the 
contributions from (i) direct $Z'$ production $Z \rightarrow q\bar{q}Z'$, 
(ii) the $Zq\bar{q}$ vertex correction, and (iii) the $Z-Z'$ mixing.  These
contributions were discussed in detail in Refs.~\cite{carmur1,carmur2},
using old LEP data, and here we simply include an updated bound.  We will
say nothing further on this point, since the corresponding exclusion
curve is superceded by the others shown in Fig.~\ref{figone}.

Other bounds on the parameter depend more crucially on the kinetic
mixing.  We consider (i) the $e^+e^-$ cross section to hadrons,
(ii) deep inelastic scattering, and (iii) the muon anomalous magnetic
moment.   In each case, however, we find that the constraints on the model 
are always weaker than those presented in Fig.~\ref{figone}. Let us 
briefly consider these topics in turn:
 
The contribution of the $Z'$ to $R = \sigma(e^+ e^- \rightarrow 
\mbox{hadrons})/
\sigma(e^+ e^- \rightarrow \mu^+ \mu^-)$ was considered in Ref.~\cite{carmur2},
and was bounded by the two-standard deviation uncertainty in the 
experimental data, using a compilation of the experimental data points.  
While this is a reasonable approximation, it does not take into 
account that a tighter bound on any new positive contribution to $R$ from 
a resonance effect is obtained when the central value of a given data point 
lies {\em below} the standard model prediction.  Here we will take this into
account, using the most precise measurements of $R$ in the $5$--$10$ GeV 
range obtained by the Crystal Ball experiment \cite{cb}.  Given the standard
model prediction for $R$ in the continuum region between the 
$J/\psi$ and the $\Upsilon$,
\begin{equation}
R=\frac{10}{3}(1+\alpha_s/\pi) \approx 3.54
\label{eq:pred}
\end{equation}
we evaluate the upper bound on the difference between theory and 
experiment taking into account two standard deviation uncertainties.
The tightest bound we obtained from any data point was $\Delta R / R < 0.05$, 
from the measurement $R=3.31\pm 0.10 \pm 0.03 \pm 0.17$ at 
$\sqrt{s}=6.25$~GeV \cite{cb}.  The first two experimental errors are 
statistical and systematic errors for the given datum, while the third is 
an  overall systematic uncertainty of $5.2\%$, which takes into account any 
average offset of the data.  Note that within the allowed parameter space of 
Fig.~\ref{figone}, $\alpha_B$ is not much larger than $10^{-2}$, and 
hence the $Z'$ width is typically of order 10 MeV, or smaller.  
On the other hand, the experimental resolution at Crystal Ball is 
$\sigma_E/E=(2.7 \pm 0.2)\% /\sqrt{E/\mbox{GeV}}$ for electromagnetically 
showering particles \cite{cb2}, so that the resolution in the $Z'$ invariant 
mass is comparable or larger to the $Z'$ width.  Assuming 
that $m_B=6.25$~GeV and $\alpha_B\approx 0.01$ (the largest value allowed 
for this mass in Fig.~\ref{figone}),  we compute the contribution to 
$\Delta R/R$ by integrating the resonant and background cross-sections over 
an energy bin equal to the detector resolution, which we set equal to 
the $Z'$ width, $\Gamma=4\alpha_B m_B/9 \approx 28$~ MeV. We find 
\[
\Delta R/R \approx 0.03
\]
which is below the experimental bound.  Since the other experimental
data points present weaker bounds on $\Delta R/R$ than the one just 
considered, we conclude that $R$ does not allow us to exclude any additional
parameter space in Fig.~1.  Note that at lower values of $\sqrt{s}$ 
above the charm threshold, $R$ is not as precisely measured, and no useful 
bounds on the model can be determined.

Deep inelastic $\nu N$ scattering,  parity violating $e N$ scattering,
and the muon $g-2$ provide only weak bounds the $Z'$ coupling.
Using the results of Ref.~\cite{carmur2}, together with the boundary
condition described earlier, we find the corresponding exclusion regions 
are given by
\begin{eqnarray}
\alpha_B & < & 0.33 (1+[m_B/\mbox{4.47 GeV}]^2)     
\,\,\,\, \mbox{ $\nu$ N scattering}  \label{eq:bnd1}\\
\alpha_B & < & 0.35 (1+[m_B/\mbox{4.47}]^2)     
\,\,\,\, \mbox{ parity-violating e N 
scattering}  \label{eq:bnd2} \\
\alpha_B & < & 1.13 (m_B/\mbox{1 GeV})^2  \,\,\,\, \mbox{ muon $g-2$}
\label{eq:mugmt}
\end{eqnarray}
which are not even visible in Fig.~\ref{figone}.  Finally, we point out
that resonant Bhabha scattering places no additional bounds on the 
model since the nonstandard contribution to the amplitude is
proportional to $c_\gamma^2$, and hence the number of events near the
resonance are suppressed relative to the electromagnetic background by
a factor of $c_\gamma^4 \sim 10^{-10}$.

Finally, we can ask how our conclusions change if the cutoff
scale $\Lambda$ is pushed to its largest possible value.  We may
use the accurate measurement the $Z$-hadronic width to first bound the mixing
parameter $c_Z(m_Z)$; we find for $\alpha_B=0.01$ that $c_Z(m_Z) 
\mbox{\raisebox{-1.0ex} {$\stackrel{\textstyle ~<~} 
{\textstyle \sim}$}} 0.02$.  This corresponds to the bound 
$\Lambda<68$~TeV.   We may obtain similar bounds from consideration of $R$; 
however these are strongly dependent on the value of $m_B$, as well as on 
the assumptions made in combining uncertainties from different, and often 
conflicting, experiments. Setting $\Lambda$ to this maximum value, we 
find $c_\gamma(m_b)\approx 0.007$, a factor of $2$ enhancement over the 
value obtained from Eq.~(\ref{eq:cees}) for the same choice of $\alpha_B$. 
Clearly, this is not significant enough to change our qualitative conclusion 
that the processes involving the kinetic mixing in 
Eqs.~(\ref{eq:bnd1}--\ref{eq:mugmt}) do little to constrain the parameter 
space of the model.

\section{Rare Decays}

What we gather from the preceding discussion is that Fig.~\ref{figone}
by itself gives a reasonable picture of the allowed parameter space
of the model.   We also learn that for $m_\psi < m_B < m_\Upsilon$
and for $m_B < m_\psi$, there are regions where the $Z'$  
coupling can be comparable to $\alpha_{EM}$.  Thus, the gauge
coupling need not be so small in these models as to require a separate leap 
of faith.  In this section, we will assume that $\alpha_B \approx \alpha$,
and consider whether the $Z'$ might eventually be detected via
rare two-body decays of charm and bottom mesons.

Since the $Z'$ coupling to fermions is purely vectorial, the Lagrangian
is charge conjugation invariant if the $Z'$ is $C$ odd.  This discrete
symmetry forbids the decays of either the $J/\psi$ or $\Upsilon$ to 
$\gamma Z'$ or $Z' Z'$ final states.  Therefore, we consider instead the 
possible two-body decays of $B$ and $D$ mesons, as well as the decays
of the lowest-lying $C$ even quarkonium states, the $\eta_c$, $\chi_c$,
$\eta_b$, and $\chi_b$.

In the first case, we know that for every $B$ or $D$ meson decay 
involving a photon in the final state, there is an analagous process 
involving the $Z'$.  The only two-body decays involving a photon
are the various $b\rightarrow s\gamma$ exclusive modes.  We
estimate
\begin{equation}
\frac{\Gamma(b\rightarrow s Z')}
{\Gamma(b\rightarrow s \gamma)} = \frac{\alpha_B}{\alpha}
(1+\frac{1}{2}\frac{m_B^2}{m_b^2}) (1-\frac{m_B^2}{m_b^2})^2
\end{equation}
where $m_b \approx 4.3$~GeV is the bottom quark mass.  While this
ratio is not necessarily small, $b\rightarrow s Z'$ is probably not the
easiest place to look for the $Z'$. Unlike $b\rightarrow s \gamma$
which is discerned experimentally by study of the photon energy spectrum,
$b\rightarrow s Z'$ yields only hadrons in the final states, and would
be overwhelmed by the larger background from 
$b\rightarrow s$~glue \cite{glue}.  On the other hand, the contribution 
to the (yet unobserved) process $b \rightarrow s e^+ e^-$ involves the 
kinetic mixing, so that for $\alpha_B \approx \alpha$ any resonance effect 
in the $e^+e^-$ invariant mass spectrum would be suppressed relative to the 
QED background by $c_\gamma^2 \sim 10^{-5}$.  The standard model prediction 
for the corresponding radiative decays in the $D$ meson system yield
drastically smaller branching fractions, and thus, these decays are not
likely to aid in the $Z'$ search.

The situation is more promising in the case of the $C$-even quarkonia states.
For example, the decay $\eta_c \rightarrow \gamma Z'$ is allowed, with
\begin{equation}
\frac{\Gamma (\eta_c \rightarrow \gamma Z')}
{\Gamma(\eta_c \rightarrow \gamma \gamma)} \approx \frac{1}{4}
\frac{\alpha_B}{\alpha} (1-m_B^2/m^2_{\eta_c})
\end{equation}
There is an overall suppression factor of $1/4$ relative to the purely 
electromagnetic decay from the squared ratio of baryon number to electric 
charge of the charm quark. In this case, one could consider 
$\eta_c \rightarrow \gamma X$, and search for a peak in the 
photon momentum spectrum.  Note that the $\eta_c$ branching fraction
to $\gamma+\mbox{hadrons}$ is dominated by the decay to $\gamma Z'$; the 
decay $\eta_c \rightarrow \gamma g$, where $g$ is a gluon, is forbidden by 
color conservation, while $\eta_c \rightarrow \gamma g g$ is forbidden by 
charge conjugation invariance.  The next possibility 
$\eta_c \rightarrow \gamma g g g$ is down by 
$\sim (\alpha_s^3/\alpha_B)/(2\pi)^4 \sim 0.001$ relative to the $Z'$ decay 
due mostly to phase space suppression, and is therefore negligible. It is 
simply an experimental question of whether single photons from 
other backgrounds processes can be adequately suppressed.  This at least seems
possible given that searches of exactly this type for lighter neutral 
gauge bosons have been undertaken in $\pi$, $\eta$ and $\eta'$ 
decays \cite{da}.  A possible scenario at an $e^+e^-$ machine is to sit 
on the $\psi (2S)$ resonance, and look for the decay 
chain 
\[
\psi (2S) \rightarrow \gamma \eta_c \rightarrow \gamma \gamma X \,\,\, .
\]  
One would retain events where one photon has precisely the right energy 
to come from the desired initial two body decay of the $\psi (2S)$, and then
study the energy spectrum of the remaining photon.    Exactly the same
procedure could be applied to $\psi (2S) \rightarrow 
\gamma \chi_c \rightarrow \gamma \gamma X$, for the various $\chi_c$ states.  
At a charm factory with a typical beam luminosity of 
$10^{34} \mbox{cm}^{-2}\mbox{sec}^{-1}$ \cite{rpp2}, and taking the 
$\psi (2S)$ production cross section to be $\sim 600$~nb from published 
data \cite{pcs}, we find $\sim 10^4$ $\gamma Z'$ events per year 
via $\chi_c$ decays, and $\sim 10^{3}$ events per year via $\eta_c$ decays.  
Here we have taken the branching fraction of the $\chi_c$ and $\eta_c$ states 
to $\gamma Z'$ to be approximately $1/4$ the $\gamma\gamma$ 
branching fractions {\em i.e.} $\sim 10^{-4}$.  The analogous decay chains of
the $\Upsilon (2S)$ in the $b$-quark system could be studied in the same
way. However, compared to the charmonium case, one would expect
a factor of $400$  reduction in the event rates: the production cross 
section for the $\Upsilon(2S)$ \cite{bpcs} is approximately two orders of 
magnitude smaller that of the $\psi (2S)$, and the $\gamma Z'$ branching 
fraction is down by a factor of $4$ relative to the same decay in the
charmonium case, due to the smaller electric charge of the $b$ quark.  Hence, 
one might still expect $\sim 25$ events/year from $\chi_b$ decays. but
the (yet unobserved) $\eta_b$ seems less promising.  
 
\section{Conclusions}

In this letter we have defined a generic class of naturally 
leptophobic $Z'$ models, and considered the $Z'$ phenomenology
in the $1$--$10$~GeV mass range, a lower range than considered
in Ref.~\cite{carmur2}.  In this mass interval, decays of various 
quarkonia states present additional bounds on the $Z'$ coupling, but 
new opportunities for its discovery as well.  We found that the experimental 
bound on $\Upsilon (1S)$ decay to two jets is primarily responsible for
defining the allowed parameter space of the model.  Bounds from
the hadronic decays of the $J/\psi$, $\psi (2S)$, $\Upsilon (2S)$,
and $\Upsilon (3S)$ only limit the parameter space in the immediate
vicinity of the resonance masses; this is a consequence of larger
experimental and (in the case of the charmonium states) theoretical
uncertainties.  We find that a $Z'$ coupling $\alpha_B \approx \alpha_{EM}$
is allowed in mass intervals above and below the charmonium threshold.
This opens the possibility of discovering the $Z'$ in rare two-body 
quarkonia decays.  We've suggested that perhaps the most interesting place to
look is in the decay chain 
$\psi (2S)\rightarrow \gamma (\eta_c\mbox{ or }\chi_c)\rightarrow \gamma 
\gamma Z'$, as well as in analogous decays of the
$\Upsilon (2S)$. If one photon has the right energy to 
indicate an initial two-body decay to the desired quarkonium state, one 
could search for a peak in the momentum distribution of the other photon.  
This could provide a stunning signal of a light and not so weakly-coupled 
$Z'$, which, given the current experimental bounds, remains a viable 
possibility.

\begin{center}              
{\bf Acknowledgments} 
\end{center}
We are grateful to David Armstrong and Carl Carlson for useful 
comments.  We thank the National Science Foundation for support under grant 
PHY-9800741.

\end{document}